\documentclass[12pt,english,final]{iopart}
\usepackage[english]{babel}
\usepackage{iopams} 
\usepackage{graphicx}
\usepackage{mathrsfs}
\usepackage{enumerate}
\usepackage{etoolbox}

\begin{document}

\def\Journal#1#2#3#4#5#6#7{#1 #2 #3 \textit{#4} \textbf{#5} #6--#7}
\def\JournalE#1#2#3#4#5#6{#1 #2 #3 \textit{#4} \textbf{#5} #6}
\def\Book#1#2#3#4#5{#1 #2 {\it #3} (#5: #4)}

\newcommand{\dd}{\mbox{d}}
\newcommand{\NN}{\mathbb{N}}
\newcommand{\ZZ}{\mathbb{Z}}
\newcommand{\RR}{\mathbb{R}}
\newcommand{\PP}{\mathbb{P}}
\newcommand{\EE}{\mathbb{E}}
\newcommand{\uu}{\mathbf{1}}
\newcommand{\RM}{\mathscr{R}}
\newcommand{\FM}{\mathscr{F}}
\newcommand{\NM}{\mathscr{N}}
\newcommand{\TT}{\mathscr{T}}
\newcommand{\step}{s}

\ifundef{\eqref}{\def\eqref#1{(\ref{#1})}}{}
\ifundef{\text}{\def\text#1{\textrm{#1}}}{}

\title[Random walk with hyperbolic probabilities]{Random walk with hyperbolic probabilities}
\author{Miquel Montero}
\address{Departament de F\'{\i}sica de la Mat\`eria Condensada,  Secci\'o de F\'{\i}sica Estad\'{\i}stica i Interdisciplin\`aria, Universitat de Barcelona (UB), Mart\'{\i} i Franqu\`es 1, E-08028 Barcelona, Spain}
\address{Institute of Complex Systems (UBICS), Universitat de Barcelona, Barcelona, Spain}
\ead{miquel.montero@ub.edu}
\vspace{10pt}
\begin{indented}
\item[]\today
\end{indented}

\begin{abstract}
The random walk with hyperbolic probabilities that we are introducing is an example of stochastic diffusion in a one-dimensional heterogeneous media. Although driven by site-dependent one-step transition probabilities, the process retains some of the features of a simple random walk, shows other traits that one would associate with a biased random walk and, at the same time, presents new properties not related with either of them. In particular, we show how the system is not fully ergodic, as not every statistic can be estimated from a single realization of the process. We further give a geometric interpretation for the origin of these irregular transition probabilities.  
\end{abstract}
\vspace{2pc}
\noindent{\it Keywords}: Random walk, Heterogeneous medium, Extreme value, Hyperbolic geometry, Ergodicity 

\submitto{J. Stat. Mech.: Theory Exp.}

\section{Introduction}

Random walks (RWs) are perhaps the most recurrent mathematical models used in the description of the erratic evolution of physical systems \cite{GW94,VK92,SR01}. In its simplest form, a RW can be thought to represent the concatenation of shifts in the position of a particle that can perform either left-ward or right-ward steps with given probabilities.  However, these one-step transition probabilities need not to be constant, they may be fully dependent on the current status of the walker and even so the stochastic process thus defined will still belong to the class of Markov chains \cite{Feller}: therefore, we can consider processes that suffer from aging \cite{EB03} or, as in the present case, particles propagating through inhomogeneous media \cite{RM01,MPW17}.

There are many mechanisms that can account for a lack of homogeneity in the one-step transition probabilities. Let us review three of them, related to the properties of the underlying medium. 
The first one is of a topological nature, and has its origin in the number of paths connecting a site: If one evenly distributes the probability of leaving any given location among its outgoing routes, and the number of possible destinies is site dependent, the one-step transition probabilities will be site dependent too. For instance, consider the case of a particle that progresses through a rooted tree with level-dependent branching order and assume that we only track  its depth within the tree structure. In such a case, the probability of returning to the parent of any given vertex is one over the degree of that vertex, different from the probability of deepening in the tree which is the number of children over the degree of the vertex. This cannot be the bare mechanism behind the lack of homogeneity proposed here since, as we will see, our one-step transition probabilities are not rational numbers. This limitation, however, can be overcome in the case of weighted networks \cite{BBPV04} where some additional feature, as e.g. channel capacity, is assigned to each (directed) link.

A second way of obtaining site-dependent transition probabilities is of geometric character: In many cases, the topological structure of the system can be embedded in a metric space directly associated, or not, with the physical 
space. This leads to the emergence of heterogeneous transition probabilities between different sites in a natural way, if one assumes that the shorter its length the greater the chances of following a given path. In particular, evidences pointing to the presence of hyperbolic geometries in technological, biological or social complex networks increase, and thus this novel topic is under very active research \cite{SKB08,MPK10,KPKVB10,AMA16,ASGB17,KKK18,GBS18}. As we show below, we do connect the properties of our heterogeneous medium with the existence of a (hidden) hyperbolic metric space, which makes our results relevant to real-world systems.     

A third method to obtain inhomogeneous transition rates is through disorder \cite{HBA02,CD18}. Quenched disorder may be the result of the interaction of an initially homogeneous medium with an external (random) potential which produces uneven transition rates between the sites \cite{DP82,CM95,CM06,SM08,SBVK10}. Particles moving across such systems usually experience anomalous diffusion \cite{MJCB14} that, ironically, in some cases can be related to the evolution of a Brownian process on the hyperbolic plane \cite{CM96,MT96,CMY98}.

A remarkable property of many (quenched and annealed) disordered systems is associated with the concept of ergodicity. A system is said to be ergodic if the ensemble average of any physical observable is equal to its time average along a single trajectory, in the limit of infinite measurement time. This means that any sample path will densely fill (except perhaps for a set of null measure) the entire phase space of the system. Therefore, strong ergodicity breaking takes place when trajectories are confined to disjoint sub-spaces. In 1992, within the context of glassy systems, Bouchaud introduced the complementary notion of weak ergodicity breaking: He considered a situation in which the system, along its evolution, gets trapped in metastable states during random periods of time, with a distribution law that leads to diverging mean sojourn times~\cite{JPB92}. 
Since then, the concept of weak ergodicity breaking has aroused great interest \cite{BB05,BB06,RB08,TS14,AB16,RDG17,SGM17,ABS18}, being commonly associated with anomalous diffusion in general, and sub-diffusion in particular. 

Here we will face a
\emph{related but different form} of non-ergodicity: In systems like the ones considered in \cite{BB05}  or \cite{BB06}, the process explores the entire phase space almost surely, but the fraction of time it spends in any given volume does not coincide with the probabilistic measure of this volume. This implies that ensemble and time averages are well defined but different. In our case, the state space is not decomposable into \emph{deterministically} inaccessible regions but into \emph{probabilistically} inaccessible regions, that is, there is always a finite probability that any given trajectory avoids a whole part of the state space, even when the observation time is infinite. 
As a consequence, the system is not ergodic since one cannot deduce all its statistical properties from a single sample path. Despite this,
the process shows selective self-averaging, depending on the physical observable considered. 

Besides that, the process has other appealing properties regarding its resemblances and differences when compared to simple and biased random walks.

The paper is structured as follows: In section \ref{Sec:process} we introduce the process at hand and derive its most basic properties. In section~\ref{Sec:expectations} we obtain explicit formulas for different expected values related to the position of the walker as means, variances and mean squared displacements. The issue of the ergodicity of  the process is first considered in this section, and rounded in section~\ref{Sec:first_times}, which is devoted to the analysis of the statistics of first-time events, as first-return probabilities or first-visit times. We provide a plausible geometric interpretation of the origin of our inhomogeneous probabilities in section~\ref{Sec:geometric}, with the help of hyperbolic metric spaces. The paper ends with section~\ref{Sec:conclusions} where conclusions are drawn and future work is envisaged.

\section{Definition and main properties of the process\label{Sec:process}}

\begin{figure}[htbp]
{
\hfill \includegraphics[width=0.8\columnwidth,keepaspectratio=true]{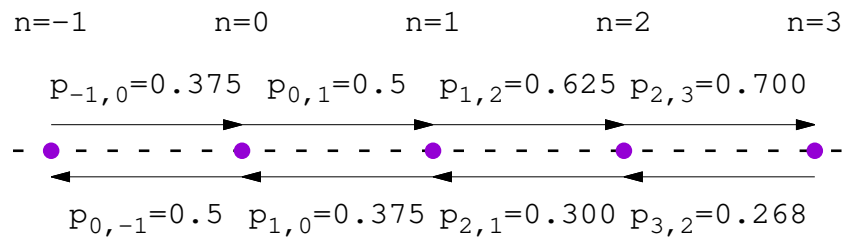}
}
\caption{
One-step transition probabilities. We show  $p_{n\to n\pm1}$ for some values of $n$, when $\xi=0.55$. Observe how $p_{n\to n\pm 1} \cdot p_{n\pm 1\to n}\approx 0.187$, independent of $n$, and that $p_{2,3}\approx0.700$ has almost attained the limiting value $\lim_{n\to \infty} p_{n\to n+ 1}=q\approx 0.750$.} 
\label{Fig:probs}
\end{figure}

Let us introduce $X_t$, the one-dimensional random walk we are going to analyze, an infinite Markov chain on the integers |namely $X_t\in\ZZ$ for $t~\in~\{0,1,2,\ldots\}$, with $X_0\equiv X_{t=0}$ given| whose one-step evolution can be expressed as follows: If at time $t$ the walker is at a given location, $X_{t}=n$, then at time $t+1$ one has
\begin{equation}
X_{t+1}=\left\{
\begin{array}{ll}
n+1,&\mbox{ with probability } p_{n\to n+1},\\
n-1,&\mbox{ with probability }  p_{n\to n-1},
\end{array}
\right.
\label{process}
\end{equation}
where we will assume the following 
dependency for the one-step transition probabilities:
\begin{equation}
p_{n\to n\pm 1}=\frac{1}{2} \frac{\cosh((n\pm1)\xi)}{\cosh (\xi) \cosh (n\xi)},
\label{eq:pn}
\end{equation} 
with $\xi\in\RR$, a parameter that controls all the transition probabilities, and whose possible origin is discussed below in section~\ref{Sec:geometric}. Although it may not be obvious at first sight, this choice for $p_{n\to n\pm 1}$ defines valid transition probabilities, i.e., they are positive and the total probability leaving any given site equals to $1$, 
\begin{equation}
p_{n\to n+1}+p_{n\to n-1}=1.
\label{eq:conservation}
\end{equation} 
Alternative expressions for \eqref{eq:pn} where the validity of equation~\eqref{eq:conservation} becomes clearer are
\begin{equation}
p_{n\to n\pm 1}=\frac{1}{2}\left[1\pm \tanh (\xi) \tanh (n \xi) \right],
\label{eq:pn2}
\end{equation} 
and
\begin{eqnarray}
p_{n\to n+ 1}&=& \frac{q^{n+1}+(1-q)^{n+1}}{q^{n}+(1-q)^{n}},\label{eq:pn3a}\\
p_{n\to n- 1}&=& q(1-q) \frac{q^{n-1}+(1-q)^{n-1}}{q^{n}+(1-q)^{n}},\label{eq:pn3b}
\end{eqnarray} 
where we have defined $q$,
\begin{equation}
q\equiv  \frac{1}{2}\left[1+\tanh (\xi)\right],
\end{equation} 
with $0< q <1$ for finite values of $\xi$. In figure~\ref{Fig:probs} we illustrate how the one-step transition probabilities vary along the lattice for a specific value of  $\xi$, $\xi=0.55$, a choice that emphasizes the idiosyncratic characteristics of the model without distorting them.

Parameter $q$ is useful for comparing $X_t$ with other well known examples of RW. The process we have defined is \emph{symmetric around the origin} with, in particular,
\begin{equation}
p_{0\to \pm 1}=\frac{1}{2},
\label{eq:p0}
\end{equation} 
for any choice of $q$. For large values of $|n|$, in turn, one finds that the one-step transition probabilities attain a limiting value. Thus, for instance, one has that
\begin{equation}
\lim_{n\to +\infty} p_{n\to n+ 1}=
\max(q,1-q),
\label{eq:limit_p}
\end{equation} 
what would define a biased random walk with positive drift, regardless of the value of $q$, $q\neq 1/2$. Note that the mirrored conclusion is obtained for negative values of $n$,
\begin{equation}
\lim_{n\to -\infty} p_{n\to n- 1}=
\max(q,1-q),
\label{eq:limit_p2}
\end{equation}
since this describes a biased random walk with \emph{negative} drift. This means that one cannot simply identify the case $q>1/2$ with a positive drift, and the case case $q<1/2$ with a negative drift: The process will behave as a simple symmetric random walk in the vicinity of the origin, and as a non-reverting, biased random walk in outer regions of the line. 
(Therefore, we could have set $\xi \geq 0$ 
without losing any generality.) 

Even though the one-step transition probabilities are inhomogeneous, the two-step ``forth-and-back" and the ``back-and-forth" probabilities do not depend on $n$:
\begin{equation}
p_{n\to n\pm 1} \cdot p_{n\pm 1\to n}=\frac{1}{4 \cosh^2 (\xi)}.
\label{eq:fnb}
\end{equation} 
This remarkable property greatly simplifies the computation of  $p_{n,t}$, the probability of finding the process at site $n$ at time $t$, if it started from the origin:
\begin{equation}
p_{n,t}\equiv \PP\left(X_t=n| X_0=0\right).
\label{eq:rho_def}
\end{equation} 
Consider the general formula
\begin{eqnarray}
p_{n,t}=\sum_{\vec{\step}\in \mathcal{S}_{n,t}} p_{0\to \step_1}  \cdots p_{ n-\step_{t}\to n},
\label{eq:rho_ini}
\end{eqnarray} 
where $\vec{\step}$  is a $t$-dimensional vector representing the actual path followed by the walker: a vector whose components are $\step_k = \pm 1$. Accordingly, $ \mathcal{S}_{n,t}$ is the subset all admissible paths given the actual initial and final locations: the set of vectors for which 
\begin{equation*}
\sum_{k=1}^{t}\step_k=n.
\end{equation*} 
The cardinality of set $\mathcal{S}_{n,t}$ is
\begin{equation}
\left|  \mathcal{S}_{n,t}\right| ={t \choose \frac{t-n}{2}}, 
\end{equation} 
if $n$ and $t$ have the same parity (i.e., if both are odd or even integers
) and zero otherwise. Please note that we are \emph{not} adopting the criterion of taking as null the binomial coefficient for non-natural arguments: We use the standard definition of the binomial coefficient based on the gamma function, 
\begin{equation}
{a \choose b}\equiv \frac{\Gamma(a+1)}{\Gamma(b+1) \Gamma(a-b+1)}, 
\end{equation} 
valid for real numbers, except negative integers $a$, $b$ or $a-b$. Therefore, a null probability is always due to the presence of an impossible event, not to the side effect of any binomial coefficient |see, e.g., equation~\eqref{eq:rho} below. 

Let us continue with the computation of \eqref{eq:rho_ini}. Assume for the moment that $n>0$, with $n\leq t$. One can easily check that
\begin{eqnarray*}
p_{0\to \step_1}  \cdots p_{ n-\step_{t}\to n}= p_{0\to 1}  \cdots p_{ n-1\to n} \cdot\frac{1}{\left[2 \cosh (\xi)\right]^{t-n}},
\end{eqnarray*}
after reordering the first occurrence of each transition probability in a directed way towards $n$ and then applying property~\eqref{eq:fnb} to the remaining $(t-n)/2$ couples. But, according to equation~\eqref{eq:pn}, one has
\begin{equation*}
p_{0\to 1}  \cdots p_{ n-1\to n}=\frac{ \cosh (n\xi)}{\left[2 \cosh (\xi)\right]^{n}},
\end{equation*}
and thus
\begin{eqnarray*}
p_{0\to \step_1}  \cdots p_{ n-\step_{t}\to n}= \frac{ \cosh (n\xi)}{\left[2 \cosh (\xi) \right]^{t}}.
\end{eqnarray*}
This quantity is the same for all the paths, and therefore
\begin{eqnarray}
p_{n,t}={t \choose \frac{t-n}{2}} \frac{ \cosh (n\xi)}{\left[2 \cosh (\xi)\right]^{t}},
\label{eq:rho}
\end{eqnarray} 
if $n$ and $t$ share the same parity, and $p_{n,t}=0$ if not. 

\begin{figure}[htbp]
{
\hfill \includegraphics[width=0.8\columnwidth,keepaspectratio=true]{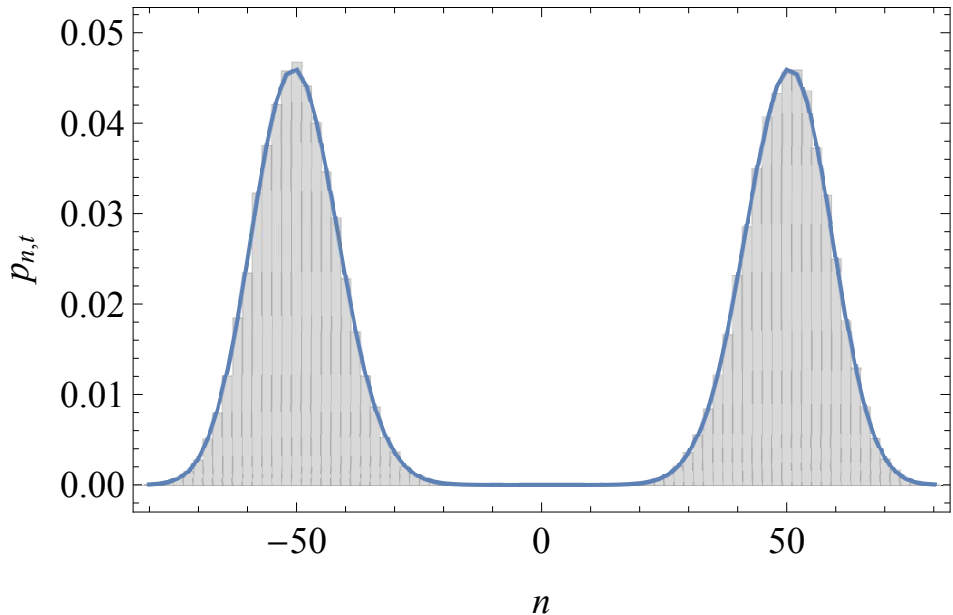}
}
\caption{
Probability function $p_{n,t}$. We depict the probability of finding the system at position $n$ after $t=100$ steps, if $X_0=0$ and $\xi=0.55$, the latter corresponding to $q\approx 0.75$.  As $t$ is an even number in this case, only even values of $n$ are shown. The solid curve is a representation of equation~\eqref{eq:rho} whereas the histogram was obtained from $100\,000$ numerical simulations of the process, with the binning (here and hereafter) chosen to include just one attainable site in each category.} 
\label{Fig:histo}
\end{figure}

Equation~\eqref{eq:rho} does not depend on the sign of $n$, while $|n|\leq t$, and thus constitutes the general solution of the stated problem. 
Note how this expression coincides with the probability function of a \emph{randomly-biased} RW,
\begin{equation}
p_{n,t}=\frac{1}{2}\left[ p^{(q)}_{n,t}+p^{(1-q)}_{n,t} \right],
\end{equation} 
with
\begin{eqnarray}
p^{(q)}_{n,t}\equiv {t \choose \frac{t-n}{2}}q^{\frac{t+n}{2}} \left(1-q\right)^{\frac{t-n}{2}},
\end{eqnarray}
since
\begin{equation}
\cosh (n\xi)=\frac{1}{2}\left[\left(\frac{q}{1-q} \right)^\frac{n}{2}+\left(\frac{1-q}{q} \right)^\frac{n}{2}\right],
\end{equation}
and
\begin{equation}
\frac{1}{\left[2 \cosh (\xi)\right]^{t}}=\left[q(1-q)\right]^\frac{t}{2}.
\end{equation}
Figure~\ref{Fig:histo} shows an example of $p_{n,t}$ for $\xi=0.55$, what corresponds to $q\approx 0.75$. 
We see how after 100 time steps, probability concentrates around $|n|\approx 50$.

The resemblance between our process and a randomly-biased RW is partial, however. Due to the lack of homogeneity, 
$X_t$ is \emph{not symmetric upon time reversal}: By following a reasoning similar to that used in the derivation of \eqref{eq:rho}, and noting that
\begin{equation*}
p_{n\to n-1}  \cdots p_{1\to 0}=\frac{1}{\left[2 \cosh (\xi)\right]^{n} \cosh (n\xi)},
\end{equation*}
one gets
\begin{equation}
\PP\left( X_{t}=0| X_0=n\right)={t \choose \frac{t-n}{2}} \frac{1}{\left[2 \cosh (\xi)\right]^{t}  \cosh (n\xi)},
\label{eq:time_rev}
\end{equation} 
which is also valid for any $n$, $|n|\leq t$, with the same parity of $t$. In fact, by using the same ideas it can be shown that one has
\begin{eqnarray}
p_{n,t;m}&\equiv&\PP\left( X_{t}=n| X_0=m\right)
={t \choose \frac{t-n+m}{2}} \frac{ \cosh (n\xi)}{\left[2 \cosh (\xi)\right]^{t}  \cosh (m\xi)},
\label{eq:tp_gen}
\end{eqnarray} 
for general values of $m$ and $n$, as long as $|n-m|\leq t$, and the involved quantities have the appropriate parity to allow reaching site $n$ starting from site $m$ in $t$ steps.~\footnote{
Note that equation~\eqref{eq:tp_gen} is sensitive to the fact that $n$ and $m$ have the same sign or not only through the binomial factor, that is, through the differing number of paths that would connect initial and final locations in each scenario, but not through the one-step probabilities themselves.}  
Equation~\eqref{eq:tp_gen} implies that $p_{n,t;m}\neq p_{m,t;n}$, unlike the case of a randomly-biased RW where
\begin{equation}
p^{(q)}_{n,t;m}= {t \choose \frac{t-n+m}{2}}q^{\frac{t+n-m}{2}} \left(1-q\right)^{\frac{t-n+m}{2}}=p^{(1-q)}_{m,t;n},
\label{eq:RBRW_time_rev}
\end{equation} 
and therefore one has
\begin{equation}
\frac{1}{2}\left[ p^{(q)}_{n,t;m}+p^{(1-q)}_{n,t;m} \right]=\frac{1}{2}\left[ p^{(q)}_{m,t;n}+p^{(1-q)}_{m,t;n} \right].
\end{equation} 
Obviously, probability $p_{n,t;m}$ in \eqref{eq:tp_gen} satisfies the corresponding Chapman-Kolmogorov equation~\cite{VK92}:
\begin{equation}
p_{n,t;m}=\sum_{n'=m-t'}^{m+t'} p_{n,t-t';n'} \cdot p_{n',t';m},
\end{equation} 
for any value of $t'$, $0< t'< t$, thanks to the fact that
\begin{eqnarray*}
\frac{ \cosh (n\xi)}{\left[2 \cosh (\xi)\right]^{t-t'}  \cosh (n' \xi)} \cdot \frac{ \cosh (n'\xi)}{\left[2 \cosh (\xi)\right]^{t'}  \cosh (m\xi)}
=\frac{ \cosh(n\xi)}{\left[2 \cosh (\xi)\right]^{t}  \cosh (m \xi)},
\end{eqnarray*} 
is independent of $n'$, and the Chu-Vandermonde identity |see, e.g.,~\cite{{HWG56}}:
\begin{equation}
{t \choose k} = 
\sum_{k'=0}^{t'}{t-t' \choose k-k'}{t' \choose k'}.
\label{eq:CV}
\end{equation} 

\section{Expectations\label{Sec:expectations}}

As the process is symmetric around the origin, the expected value of the position is zero if the process began at that point, $\EE\left[ X_t|X_0=0\right]=0$. The situation changes drastically when one considers a general initial point, since then
\begin{eqnarray}
\mu_{t;m}&\equiv&\EE\left[X_t|X_0=m\right]=\sum_{n=m-t}^{m+t}n\cdot  p_{n,t;m}\nonumber\\
&=&\sum_{k=0}^{t} {t \choose k} \frac{m+2 k -t}{ \cosh (m\xi)} \frac{ \cosh ((m+2 k -t)\xi)}{\left[2 \cosh (\xi)\right]^{t} }\nonumber\\
&=&m+\tanh (\xi) \tanh (m\xi) \cdot t,
\label{eq:ep_gen}
\end{eqnarray} 
and the process shows ballistic behavior for $\xi\neq0$, despite the existing symmetry \cite{GNSS12}. This result is very relevant, since the process is time homogeneous and then
\begin{equation*}
\EE\left[X_{t+\tau}|X_{t}=m\right]=m+\tanh (\xi) \tanh (m\xi) \cdot \tau=\mu_{\tau;m},
\end{equation*}
for any $t, \tau\geq 0$. 
In addition, note how we can rewrite $\mu_{t;m}$ as
\begin{equation}
\mu_{t;m}=m+V_m\cdot t,
\label{eq:mean} 
\end{equation}
where we have introduced the current of  probability leaving site $m$, $V_m$, 
\begin{equation}
V_m\equiv p_{m\to m+1} - p_{m\to m-1}=\tanh (\xi) \tanh (m \xi),
\end{equation}
a quantity that, as the symbol suggest, plays the role of the \emph{initial position-dependent velocity} of the process starting from $m$. 
This speed increases with $|m|$, until reaching a terminal value: 
\begin{equation}
\lim_{m\to \pm \infty} V_{m}=\pm |\tanh (\xi)|.
\label{eq:terminal_velocity}
\end{equation}

\begin{figure}[htbp]
{

\hfill (a) \includegraphics[width=0.37\columnwidth,keepaspectratio=true]{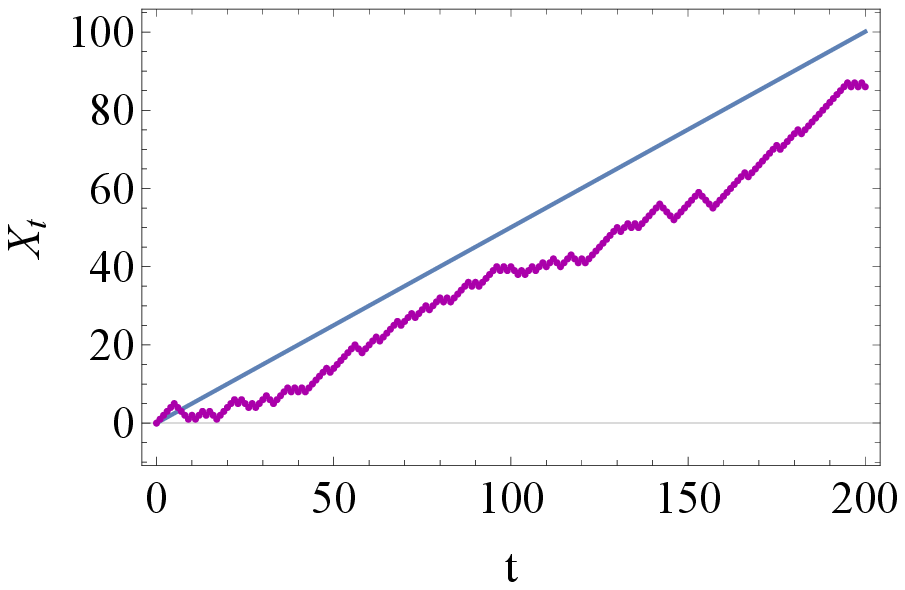} (b) \includegraphics[width=0.37\columnwidth,keepaspectratio=true]{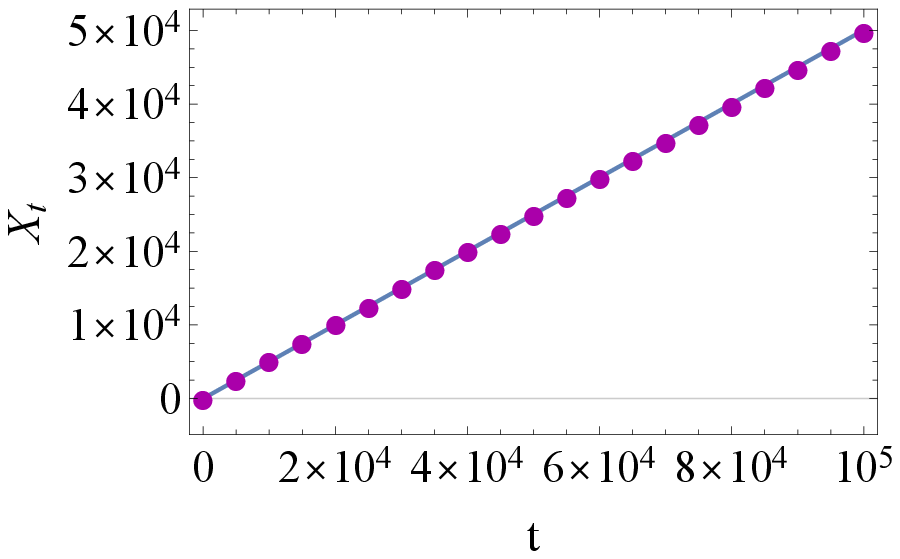}
}
\caption{
Sample path of $X_t$, for $\xi=0.55$. (a) In the left panel we illustrate the beginning of the simulation of a realization of the process with $100\,000$ time steps. (b) In the right panel we consider the whole evolution of the walker, by plotting its position every $5\,000$ time steps, for clarity reasons. 
In both plots, the solid line has a slope equal to the terminal velocity corresponding to positive walks, see equation~\eqref{eq:terminal_velocity}.
} 
\label{Fig:sample_path}
\end{figure}

In contrast to the linear growth in $t$ of $\mu_{t;m}$ for $m\neq 0$, the time evolution of the standard deviation of the process, $\sigma_{t;m}$, presents two well-different regimes:  
\begin{eqnarray}
\sigma^2_{t;m}&\equiv&\EE\left[X^2_t|X_0=m\right]-\mu^2_{t;m}
=\sum_{n=m-t}^{m+t}n^2\cdot  p_{n,t;m}-\mu^2_{t;m}\nonumber\\
&=&\sum_{k=0}^{t} {t \choose k} \frac{(m+2 k -t)^2}{ \cosh (m\xi)} \frac{ \cosh ((m+2 k -t)\xi)}{\left[2 \cosh (\xi)\right]^{t} }-\mu^2_{t;m}\nonumber\\
&=&m^2+2 m V_m  t +\frac{1}{\cosh^2(\xi)} t + \tanh^2 (\xi) t^2-\mu^2_{t;m}\nonumber\\
&=&\frac{1}{\cosh^2(\xi)} t+ \frac{\tanh^2(\xi)}{\cosh^2 (m\xi)}t^2.
\label{eq:sigma_gen}
\end{eqnarray}  
The first term in the last expression will be the most relevant at the beginning of the evolution if either $|\xi|$ is small 
or $|m|$ is large but, as time increases, the second term will eventually dominate.  Indeed one has 
\begin{equation}
\lim_{t\to \infty} \frac{\mu_{t;m}}{\sigma_{t;m}}=|\sinh (m\xi)|,
\end{equation}
for fixed $m$, and
 \begin{equation}
\lim_{|m| \to \infty} \frac{ |\mu_{t;m}-m|}{\sigma_{t;m}}=|\sinh (\xi)| \cdot\sqrt{t},
\end{equation}
for fixed $t$. This translates into trajectories that show a marked directionality and a relatively low dispersion, as it can be observed in figure~\ref{Fig:sample_path}. Actually, the example displayed here corresponds to a case in which, starting from $X_0=0$, $X_t$ does not take negative values at any moment. We defer until section~\ref{Sec:first_times} the proof that this fact is neither incidental nor a consequence of considering a sample path with finite length.

Having a finite 
probability of obtaining a trajectory that avoids half of the space of states 
(when $\xi\neq 0$) prevents the process from being ergodic: For instance, one will not be able to estimate 
$p_{n,t}$ or $\mu_{t;n}$ for arbitrary choices of $n$ from that sample path. Despite this, the process is 
partially ergodic: There are magnitudes that \emph{can} be evaluated accurately with the knowledge of a single trajectory, regardless of its peculiarities, being paradoxically one of these quantities the mean squared displacement (MSD), a statistic that is commonly used to verify the ergodicity of a process~\cite{MJCB14,TS14,SGM17}. 

To prove this, let us define the squared displacement of the process between times $t$ and $t+\tau$, $\tau\geq 0$, $\Delta X^2(t,t+\tau)$, as
\begin{equation}
\Delta X^2(t,t+\tau)\equiv  (X_{t+\tau}-X_{t})^2.
\label{eq:MSD}
\end{equation}
The time-averaged MSD, $\overline{\Delta X^2(\tau;T)}$, is the normalized sum of the all the different squared displacements that can be obtained from a single realization along the measurement time $T$, $T\geq \tau$,
\begin{equation}
\overline{\Delta X^2(\tau;T)} \equiv\frac{1}{T-\tau+1} \sum_{t=0}^{T-\tau}\Delta X^2(t,t+\tau).
\label{eq:TAMSD}
\end{equation}
Let us compute next the expected value of the squared displacement, the MSD,
\begin{eqnarray}
\EE\left[\Delta X^2(t,t+\tau)|X_0=0\right]&=&
\EE\left[X^2_{t+\tau}|X_0=0\right]+  \EE\left[X^2_{t}|X_0=0\right] \nonumber\\
&-&2 \EE\left[X_{t+\tau}X_{t}|X_0=0\right].
\label{eq:EMSD}
\end{eqnarray}
We already have expressions for the first two terms, since $\EE\left[X^2_{t+\tau}|X_0=0\right]=\sigma^2_{t+\tau;0}$ and  $\EE\left[X^2_{t}|X_0=0\right]=\sigma^2_{t;0}$, so we can focus our efforts in the last one:
\begin{equation}
\EE\left[X_{t+\tau}X_{t}|X_0=0\right]=\EE\left[\left. \EE\left[X_{t+\tau}X_{t}|X_{t}\right]\right|X_0=0\right],
\end{equation}
by virtue of the tower property of the conditional expectation, but
\begin{equation}
\EE\left[X_{t+\tau}X_{t}|X_{t}=m\right]=\mu_{\tau;m} \cdot m=X_{t}^2+ X_{t} \tanh (\xi) \tanh(X_{t}\xi)\cdot \tau,
\end{equation}
and therefore
\begin{eqnarray}
\EE\left[X_{t+\tau}X_{t}|X_0=0\right]&=&\EE\left[ X^2_t|X_0=0\right]+\tau \tanh(\xi) \EE\left[X_{t}  \tanh(X_{t}\xi)|X_0=0\right]\nonumber \\
&=& \sigma^2_{t;0}+\tau \tanh (\xi)  \sum_{n=-t}^{t} n  \tanh \left(n \xi\right) p_{n,t}\nonumber \\
&=& \sigma^2_{t;0}+\tau \tanh (\xi) \sum_{k=0}^{t} {t \choose k} (2 k -t)  \frac{ \sinh ((2 k -t)\xi)}{\left[2 \cosh (\xi)\right]^{t} }\nonumber \\
&=& \sigma^2_{t;0}+\tau  t \tanh^2 (\xi).
\end{eqnarray}
Collecting all the pieces, we finally get
\begin{eqnarray}
\EE\left[\Delta X^2(t,t+\tau)|X_0=0\right]&=&
\sigma^2_{t+\tau;0}+ \sigma^2_{t;0}-  2  \sigma^2_{t;0} -2 \tau  t \tanh^2 (\xi)\nonumber\\
&=& \frac{1}{\cosh^2(\xi)} \tau+ \tanh^2(\xi) \tau^2\equiv\sigma^2_{\tau;0},
\label{eq:EMSD_final}
\end{eqnarray}
independent of $t$.~\footnote{After a similar although more involved calculation, it can be shown that $\EE\left[\Delta X^2(t,t+\tau)|X_0=m\right]$ is not a function of $m$ either, that is $\EE\left[\Delta X^2(t,t+\tau)|X_0=m\right]=\sigma^2_{\tau;0}$, for any value of $m$.} Therefore, one has
\begin{equation}
\EE\left[\left.\overline{\Delta X^2(\tau;T)}\right|X_0=0\right]= \sigma^2_{\tau;0},
\label{eq:ETAMSD}
\end{equation}
independent of $T$. 

\begin{figure}[htbp]
{

\hfill (a) \includegraphics[width=0.37\columnwidth,keepaspectratio=true]{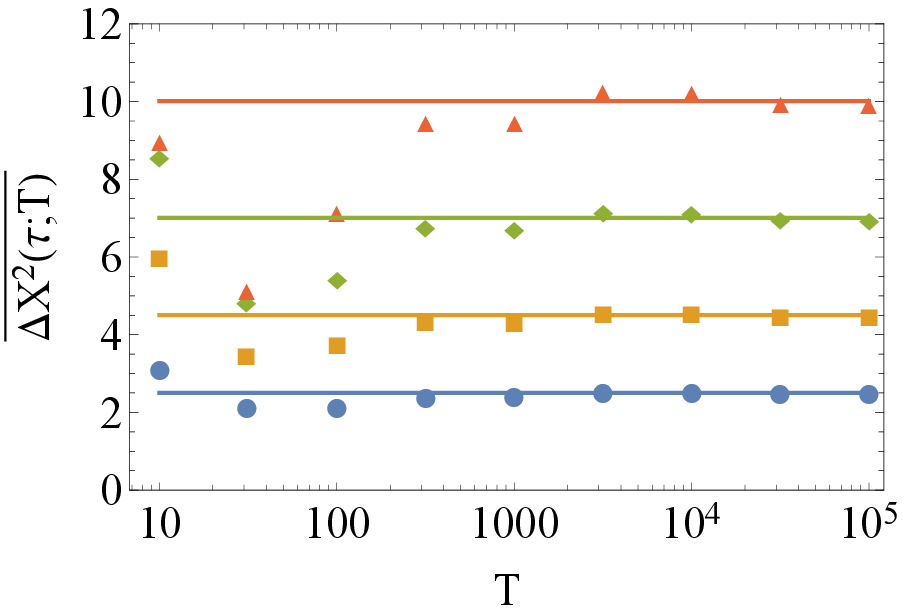} (b) \includegraphics[width=0.37\columnwidth,keepaspectratio=true]{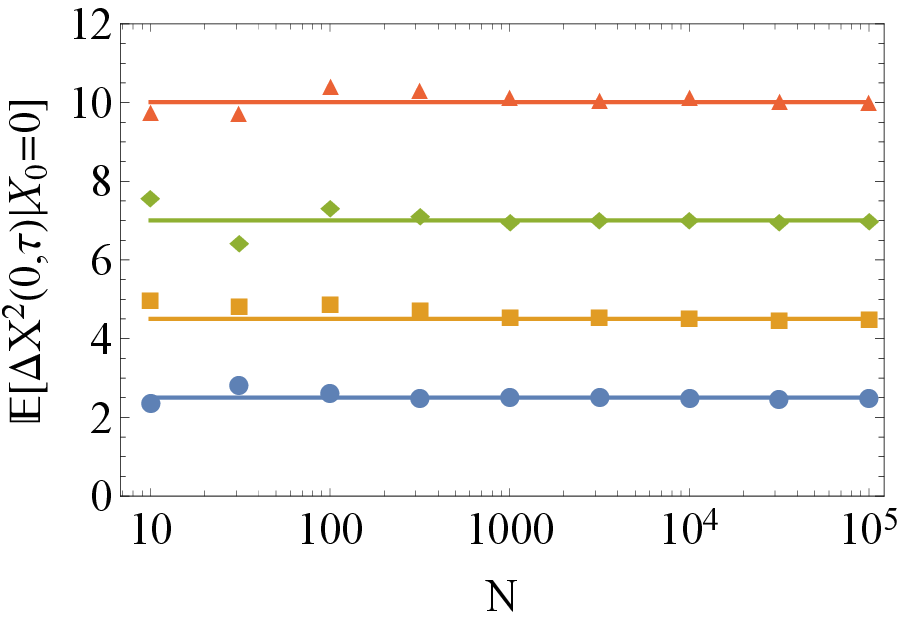}

}
\caption{
Mean squared displacement of $X_t$, for $\xi=0.55$ and $X_0=0$. (a) In the left panel we show the time-averaged MSD, $\overline{\Delta X^2(\tau;T)}$, for $\tau$ equal to 2  (blue dots), 3 (orange squares), 4 (green diamonds), and 5 (red triangles). The dataset is the same used in the confection of figure~\ref{Fig:sample_path}, a single sample path of $100\,000$ time steps. 
The value is plotted against the observation time $T$. (b) In the right panel we consider the 
ensemble estimation of MSD, $\EE\left[\Delta X^2(0,\tau)|X_0=0\right]$. The dataset is the same used in the confection of figure~\ref{Fig:histo},  $100\,000$ trajectories of $100$ time steps each. The estimate is plotted against  $N$, the number of samples averaged. Only the initial displacement is considered. In both graphs, the solid lines correspond to the theoretical values of $ \sigma^2_{\tau;0}$.
}
\label{Fig:MSD}
\end{figure}

In figure~\ref{Fig:MSD}.(a) we can observe how $\overline{\Delta X^2(\tau;T)}$ converges to this theoretical value as $T$ increases, for different choices of $\tau$, $\tau=2,3,4$ and 5.~\footnote{The case $\tau=1$ leads to $\Delta X^2(t,t+1)=1$, identically.} We have used the same sample path of $100\,000$ time steps of figure~\ref{Fig:sample_path}. The relative error reduces as the number of terms in~\eqref{eq:TAMSD} increases, i.e., is $O\left(T^{-1/2}\right)$, for $T\gg \tau$.  
For comparison purposes, we have included in the same figure,  figure~\ref{Fig:MSD}.(b), a second graph where the MSD, $\EE\left[\Delta X^2(0,\tau)|X_0=0\right]$, is estimated from the initial displacements of $N$ instances of the $100\,000$ replicas of the process used in the making of figure~\ref{Fig:histo}. 
In this case the convergence scales as $O\left(N^{-1/2}\right)$, as the number of realizations considered grows. 

\section{First-time events\label{Sec:first_times}}

It is clear from the structure of equation~\eqref{eq:tp_gen} that the transition probability it is not invariant under a spatial shift. Despite that, the probability that the process returns to any given point $n$ after $2t$ steps, $u_{2t}$,
\begin{equation}
u_{2t}\equiv\PP\left(X_{2t}=n| X_0=n\right)= {2t \choose t} \frac{1}{\left[2 \cosh (\xi)\right]^{2t}},
\label{eq:return}
\end{equation} 
does not depend on $n$, and coincides with the expression that one would associate with the simple random walk, after the replacement of the fair-coin toss probability $1/2$ by the factor  $1/2  \cosh (\xi)$. Equation~\eqref{eq:return} allows us to check that $p_{n,t;m}$ |cf. equation~\eqref{eq:tp_gen}| fulfills the right closure property: e.g., consider the case
\begin{eqnarray}
u_{2t}&=&\PP\left(X_{2t}=0| X_0=0\right)\nonumber \\
&=& \sum_{n=-t}^{t} \PP\left(X_{2t}=0| X_t=n\right)\PP\left( X_t=n| X_0=0\right)\nonumber\\
&=& \sum_{n=-t}^{t} {t \choose \frac{t-n}{2}}^2 \frac{1}{\left[2 \cosh (\xi)\right]^{2t}}= {2t \choose t} \frac{1}{\left[2 \cosh (\xi)\right]^{2t}},
\end{eqnarray} 
where we have applied again the Chu-Vandermonde identity, equation~\eqref{eq:CV}, having in mind that the last sum only contains those terms for which $n$ has the same parity of $t$.

A quantity related to $u_{2t}$ is $f_{2t}$, the probability that the process returns to a given point $n$ after $2t$ steps \emph{for the first time}. Before analyzing $f_{2t}$, let us introduce the random variable $\TT_{n;m}$,
 \begin{equation}
\TT_{n;m}\equiv\min\left\{ t>0 : X_{t}=n| X_{0}=m\right\},
\label{eq:T_gen_def}
\end{equation}
i.e., the interval defined by the first-visit time to an arbitrary site $n$ starting from $m$~\cite{SR01}. Now, we can write $f_{2t}$ in terms of $\TT_{n;m}$,
\begin{equation}
f_{2t}\equiv \PP\left( \TT_{n;n}=2t\right).
\label{eq:frt}
\end{equation} 
As the notation suggest, $f_{2t}$ is independent of $n$, just like $u_{2t}$. In fact, $f_{2t}$ can be related to $u_{2t}$ by the same formula that one has for the simple random walk:
\begin{equation}
f_{2t}=\frac{1}{2t-1} u_{2t}=\frac{1}{2t-1} {2t \choose t} \frac{1}{\left[2 \cosh (\xi)\right]^{2t}},
\end{equation} 
since in both cases $f_{2t}$ and $u_{2t}$ do satisfy the following relationship:
\begin{equation}
u_{2t}=\sum_{t'=1}^{t} f_{2t'}\cdot u_{2(t-t')}.
\end{equation} 

\begin{figure}[htbp]
{
\hfill \includegraphics[width=0.8\columnwidth,keepaspectratio=true]{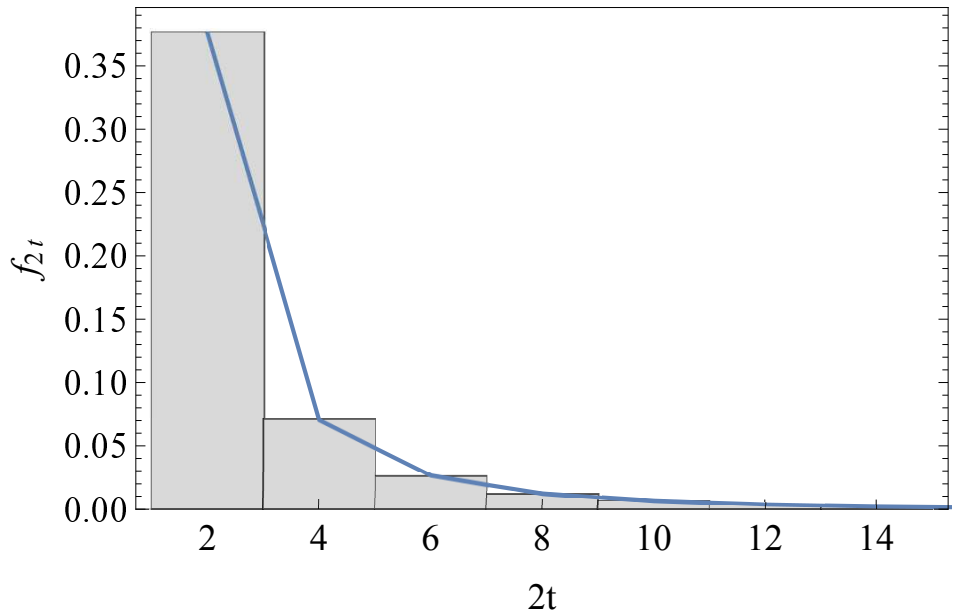}
}
\caption{
First-visit probability function $f_{2t}$. In this graphic we plot the theoretical value for the probability that the first return to the initial location occurred after $2t$ steps, solid curve, together with an estimation obtained from the first-return time lag to every sample $X_t$ of our single trajectory of $100\,000$ time steps. Note that, indeed, the return only took place in $50\,184$ of the cases, very close to the theoretical prediction of eventually returning to the starting point of $0.49948$ of the cases, see equation~\eref{eq:transient}.
} 
\label{Fig:f2t}
\end{figure}

This independency of $f_{2t}$ with respect to $n$ can be used to illustrate another instance of the 
partial ergodicity shown by the process. The time homogeneity determines that $\TT_{n;n}$ can be also defined as
\begin{equation}
\TT_{n;n}\equiv\min\left\{ \tau>0 : X_{t+\tau}=n| X_{t}=n\right\},
\label{eq:T_gen_def_alt}
\end{equation}
and since $\PP\left( \TT_{n;n}=2t\right)$ is not a function of $n$, a single trajectory can be used to estimate $f_{2t}$, as we have done in figure~\ref{Fig:f2t} where the successive values of $X_t$ are taken from our realization of the process with $100\,000$ time steps.

Note that when obtaining this estimation of $f_{2t}$, we must not discard those points in the trajectory for which no first-visit time lapse can be defined, thus reducing the total amount of events. The reason is because the process (for $\xi \neq 0$) is \emph{not recurrent}:
\begin{eqnarray}
\PP\left(\exists t>0:  X_{2t}= n| X_0=n \right)
&=&\sum_{t=1}^{\infty} f_{2t}=1-|\tanh (\xi)|
<1.
\label{eq:transient}
\end{eqnarray} 
Even being the process transient, 
the mean number of returns to any given point may be still greater that one,
\begin{equation}
\sum_{t=1}^{\infty} u_{2t} = \sum_{t=1}^{\infty} {2t \choose t} \frac{1}{\left[2 \cosh (\xi)\right]^{2t}} =|\coth (\xi)|-1\geq 1,
\label{eq:mnr}
\end{equation} 
which is true for $|\xi| \leq 0.5 \ln 3 \approx 0.55$, very similar to the value we are using in the illustrative examples. 
Indeed, the mean number of returns grows unboundedly as $\xi \to 0$.

\begin{figure}[htbp]
{
\hfill \includegraphics[width=0.8\columnwidth,keepaspectratio=true]{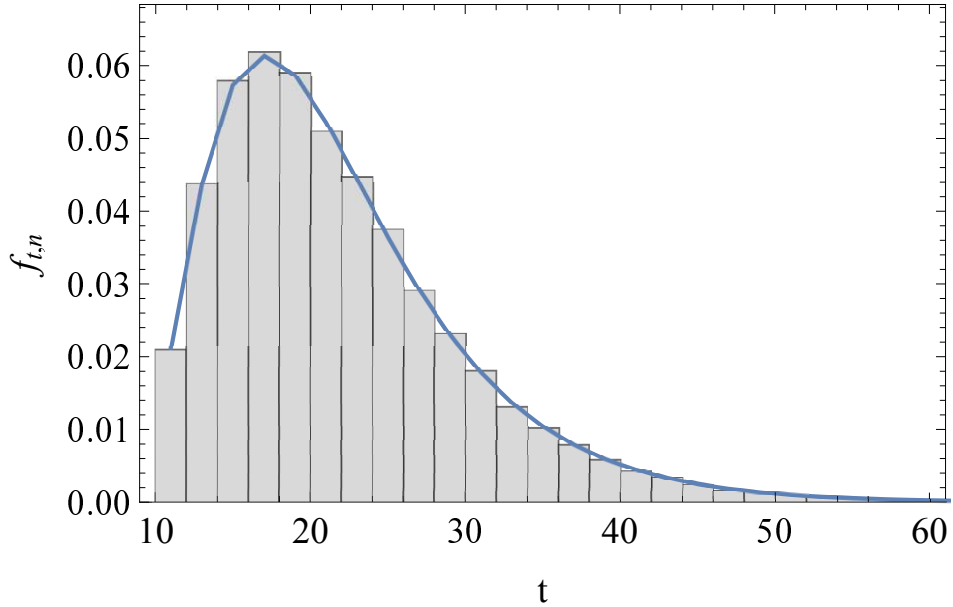}
}
\caption{
First-visit probability function $f_{t,n}$. In this graphic we plot the probability that the first visit at site $n=11$  occurred after $t$ steps, for $X_0=0$, $\xi=0.55$.  As $n$ is an odd number, only those odd values of $t$ equal or above 11 are considered. The solid curve is a representation of equation~\eqref{eq:ftn} whereas the histogram was obtained from the same $100\,000$  simulated trajectories used in the confection of figure~\ref{Fig:histo}, where $t_\text{max}=100$. In this case, only $50\,099$ of the sample paths have ever reached $n=11$, in excellent agreement with the theoretical prediction of about $50\,000$ for $t_\text{max} \to \infty$, $\PP\left(\RM_{11;0}\right)\approx 0.500003$, see equation~\eqref{eq:PTT}.
} 
\label{Fig:ftn}
\end{figure}

Consider now the problem of finding the statistics of the first-visit time to an arbitrary site $n$ starting from the origin, $f_{t,n}$,
\begin{equation}
f_{t,n}\equiv \PP\left( \TT_{n;0}=t\right),
\label{eq:fpt}
\end{equation} 
which is null if $n$ and $t$ have different parity, or $t<|n|$. For $n=0$ one has that $f_{t,0}=f_{2t}$, whereas for $n>0$, we can obtain $f_{t,n}$ thanks to its connection with $p_{n,t}$ and $u_{t}$,
\begin{equation}
p_{n,t}= 
\sum_{t'=n}^{t} f_{t',n} \cdot u_{t-t'},
\label{eq:fpt_closure}
\end{equation} 
since the probability of finding the process at site $n$ at time $t$ is the sum of all paths that reach $n$ at time $t'\leq t$, and then return to $n$ in the remaining time span. Since $u_{t-t'}$ is equal to zero for $t'-t$ odd, the three integers $n$, $t$ and $t'$ must share the same parity. With this proviso, let us introduce the $z$-transform of $p_{n,t}$, $\hat{p}_{n}(z)$,~\footnote{Depending on the context, alternative definitions of this discrete version of the Laplace transform can be found in the literature.} 
\begin{equation}
\hat{p}_{n}(z) \equiv\sum_{t=n}^{\infty} p_{n,t}\cdot z^t= \sum_{t=n}^{\infty} {t \choose \frac{t-n}{2}} \frac{  \cosh (n\xi)}{\left[2 \cosh (\xi)\right]^{t}}\cdot z^t,
\label{eq:mgf}
\end{equation}
with $0\leq z< \cosh (\xi)$, and consider our previous findings, equations~\eqref{eq:return} and~\eqref{eq:fpt_closure},
\begin{eqnarray*}
\hat{p}_{n}(z) &=& \sum_{t=n}^{\infty} \sum_{t'=n}^{t}  f_{t',n} {t-t' \choose \frac{t-t'}{2}} \frac{1}{\left[2 \cosh (\xi)\right]^{t-t'}} \cdot z^t\\
&=& \sum_{t'=n}^{\infty}  f_{t',n} \cdot z^{t'}\sum_{t=t'}^{\infty} {t-t' \choose \frac{t-t'}{2}} \cdot \mathcal{Z}^{t-t'},
\end{eqnarray*} 
where
\begin{equation}
\mathcal{Z} \equiv \frac{z}{2 \cosh (\xi)}<\frac{1}{2}.
\end{equation}
But
\begin{eqnarray*}
\sum_{t=t'}^{\infty} {t-t' \choose \frac{t-t'}{2}} \cdot \mathcal{Z}^{t-t'}=\sum_{k=0}^{\infty} {2 k \choose k} \cdot \mathcal{Z}^{2 k}=\frac{1}{\sqrt{1-4  \mathcal{Z}^{2}}},
\end{eqnarray*} 
what renders
\begin{eqnarray*}
 \sum_{t'=n}^{\infty}  f_{t',n} \cdot z^{t'} &=&\hat{p}_{n}(z)  \sqrt{1-4  \mathcal{Z}^{2}}
 =\hat{p}_{n}(z) \sum_{k=0}^{\infty} (-1)^k {1/2 \choose k} \frac{z^{2 k}}{\cosh^{2 k} (\xi)}.
\end{eqnarray*}
Therefore, from equation~\eqref{eq:mgf}, one gets~\footnote{Observe how this result is not valid for $n=0$ and becomes ill-defined for $n=t=0$.}
\begin{eqnarray}
f_{t,n} &=&\frac{  \cosh (n\xi)}{\left[2 \cosh (\xi)\right]^{t}} \sum_{k=0}^{\frac{t-n}{2}} {t -2 k\choose \frac{t-n}{2}-k}  (-1)^k {1/2 \choose k} 2^{2 k}\nonumber\\
&=&\frac{  \cosh (n\xi)}{\left[2 \cosh (\xi)\right]^{t}} {t \choose \frac{t-n}{2}}\frac{n}{t}=\frac{n}{t}  p_{n,t},
\label{eq:ftn}
\end{eqnarray}
once again, a result that is formally identical to the one corresponding to a simple random walk. For negative targets, $n$ must be replaced by $|n|$, since $p_{n,t}$ is already symmetric. We show in figure~\ref{Fig:ftn} the shape of probability $f_{t,n}$ for $n=11$. Obviously, the minimum feasible value of $t$ is $t=11$, although the probability attains a maximum at $t=17$.

Equation~\eqref{eq:ftn} can be easily generalized to include an arbitrary choice for the initial starting site $m$, $m\neq n$: 
\begin{eqnarray}
f_{t,n;m}&\equiv&
\PP\left( \TT_{n;m}=t\right)
=\frac{|n-m|}{t}p_{t,n;m}\nonumber \\
&=&\frac{|n-m|}{t}{t \choose \frac{t-n+m}{2}} \frac{ \cosh (n\xi)}{\left[2 \cosh (\xi)\right]^{t}  \cosh (m\xi)},
\label{eq:ftnm}
\end{eqnarray}
where the same restrictions to values and parities of $n$, $m$ and $t$ considered in the derivation of expression~\eqref{eq:tp_gen} do apply here.

Let us prove now this announced lack of ergodicity in the process. 
To that end, we will define $\RM_{n;m}$ as the set of all the paths that go from site $m$ to site $n$: 
 \begin{equation}
\RM_{n;m}\equiv \left\{\exists t>0 : X_{t}=n| X_{0}=m\right\}.
\label{eq:R_gen_def}
\end{equation}
Then, for $\xi\neq 0$, the probability that any given site $n$, starting from $m$, is ever reached is \emph{always} smaller than one and amounts to
\begin{eqnarray}
\PP\left(\RM_{n;m}\right)&=& \sum_{t=|n-m|}^{\infty} f_{t,n;m} =  \sum_{t=|n-m|}^{\infty} \frac{|n-m|}{t}p_{t,n;m}\nonumber \\ 
&=&{}_2F_1\left(\frac{\ell}{2},\frac{\ell+1}{2};\ell+1;\frac{1}{\cosh^2 (\xi)}\right)\frac{\cosh (n\xi)}{ \cosh (m \xi)  \left[2 \cosh (\xi)\right]^{\ell} },
\label{eq:PTT}
\end{eqnarray} 
where $\ell\equiv |n-m|>0$ and ${}_2F_1\left(a,b;c;u\right)$ is the Gaussian hypergeometric function. Recall that equation~\eqref{eq:PTT} does not hold when $n=m$. This case must be considered separately, as we have already done, c.f. equation~\eqref{eq:transient},
\begin{equation*}
\PP\left(\RM_{n;n} \right)= 1-|\tanh (\xi)|<1.
\end{equation*}
Therefore, the process is not ergodic: For any starting point of the process, the set of random trajectories that do not fill the entire state space is not of null measure. In particular, consider the case of $\PP\left(\RM_{\pm1;0}\right)$, 
\begin{eqnarray}
\PP\left(\RM_{\pm 1;0}\right)
&=& {}_2F_1\left(\frac{1}{2},1;2;\frac{1}{\cosh^2 (\xi)}\right)\frac{1}{2}=\frac{1+e^{-2|\xi|}}{2}.
\end{eqnarray} 
This result implies that the chances that a whole path is restricted either to non-negative integers or non-positive integers are the same:
\begin{eqnarray}
\PP\left( X_{t}\geq 0, \forall t\geq 0 | X_0=0 \right)&=&\PP\left( X_{t}\leq 0, \forall t\geq0| X_0=0 \right)=\frac{1-e^{-2|\xi|}}{2},
\end{eqnarray}
whereas the probability that in a single realization one finds visits to both positive and negative sites reduces exponentially with $|\xi|$,
\begin{eqnarray}
\PP\left(\exists \{t'\geq 0, t''\geq 0\} : X_{t'} \cdot X_{t''} < 0 | X_0=0 \right)=e^{-2 |\xi|}.
\end{eqnarray}
For $\xi=0.55$, these three probabilities are roughly equal, as it can be checked in table~\ref{table:PR}. Therefore, a sample path like the one in figure~\ref{Fig:sample_path} is more the rule than the exception under these circumstances.

\begin{table}[htbp]
\caption{\label{table:PR}  
Probabilities that a realization of $X_t$ is restricted to non-negative integers, to non-positive integers, and that it takes positive and negative values. We show theoretical values for $\xi=0.55$, along with estimates obtained from the analysis of the $N=100\,000$ trajectories with $t_\text{max}=100$ used in the confection of figures~\ref{Fig:histo} and~\ref{Fig:ftn}. The relative error is of $O\left(N^{-1/2}\right)$ again.
}
\begin{indented}
\item[]\begin{tabular}{@{}lll}
\br
\hfill& Theoretical & Estimate\\
\mr
$\PP\left( X_{t}\geq 0, \forall t\geq 0 | X_0=0 \right) $  & 0.33356 & 0.33533\\
$\PP\left( X_{t}\leq 0, \forall t\geq0| X_0=0 \right)$  & 0.33356 & 0.33307\\
$ \PP\left(\exists  \{t'\geq 0, t''\geq 0\} : X_{t'} \cdot X_{t''} < 0 | X_0=0 \right) $ & 0.33287 & 0.33160\\
\br
\end{tabular}
\end{indented}
\end{table}  

As a final curiosity, the mean first-visit time, provided that the time lapse is finite, does exist and has a surprisingly compact expression: 
\begin{eqnarray}
\EE\left(\left.\TT_{n;m}\right|\RM_{n;m} \right)&\equiv&\frac{ \displaystyle\sum_{t=|n-m|}^{\infty} t f_{t,n;m}}{ 
\PP\left(\RM_{n;m}\right)}
=\frac{{}_2F_1\left(\frac{\ell+1}{2},\frac{\ell+2}{2};\ell+1;\frac{1}{\cosh^2 (\xi)}\right)}{{}_2F_1\left(\frac{\ell}{2},\frac{\ell+1}{2};\ell+1;\frac{1}{\cosh^2 (\xi)}\right)}\ell  \nonumber \\
&=&\left|\coth (\xi)\right| \ell,
\end{eqnarray} 
for $m \neq n$, and 
\begin{eqnarray}
\EE\left(\left.\TT_{n;n}\right|\RM_{n;n} \right)&=&\frac{ \displaystyle\sum_{t=1}^{\infty} 2 t f_{2t}}{ \PP\left(\RM_{n;n} \right)}=\frac{1}{\cosh(\xi)|\sinh(\xi)| (1-|\tanh (\xi)|)}\nonumber\\
&=&|\coth (\xi)|+1,
\end{eqnarray} 
for $m=n$. Therefore, we have found a translational invariance that is not present in $f_{t,n;m}$, see equations~\eqref{eq:ftnm} and~\eqref{eq:PTT}. Another aspect of the 
partial ergodicity shown by the process.

\section{A geometric interpretation of one-step probabilities\label{Sec:geometric}}

We discuss now one possible mechanism for generating the inhomogeneous one-step probabilities that drive the dynamics of $X_t$. The approach is based on the idea that coexisting with the topological structure of the state space of the process, $\ZZ$ in our case, there is an underlying metric space that assigns a distance $d_{n,m}$ to integers $n$ and $m$, different from the $L_1$ distance $|n-m|$, and that this distance determines the likelihood of the transitions on the basis of the relative proximities of the possibles destinies to the actual location of the walker. 

So that, let us introduce $d_{n,m}$,
\begin{equation}
d_{n,m}\equiv \frac{\sinh(|n-m| \xi)}{\tanh (\xi) \cosh (n\xi) \cosh (m\xi)}.
\label{eq:dist}
\end{equation}
Equation~\eqref{eq:dist} defines a distance between sites $m$ and $n$ as it shows all the required properties: it is semi-positive definite (for any value of $\xi$), symmetric, gives a null result if and only if $n=m$, and fulfills the triangle inequality
\begin{equation}
d_{n,m}\leq d_{n,l}+d_{l,m},
\end{equation}
since
\begin{equation}
d_{n-1,n+1}= d_{n-1,n}+d_{n,n+1}.
\end{equation}
Then, if we assume that the probability of a one-step transition is inversely proportional to the distance, i.e.,
\begin{equation}
\frac{p_{n\to n+1}}{p_{n\to n-1}}=\frac{d_{n-1,n}}{d_{n,n+1}}=\frac{\cosh ((n+1)\xi)}{\cosh ((n-1)\xi)},
\label{eq:prob_distance}
\end{equation}
one obtains equation~\eqref{eq:pn} after demanding the conservation of probability, equation~\eqref{eq:conservation}. Note that the factor $\tanh(\xi)$ of the denominator in equation~\eqref{eq:dist} is discretional, in the sense that it has no effect in equation~\eqref{eq:prob_distance} and consequently does not interfere in the computation of the one-step transition probabilities. However, it provides several interesting features: It makes $d_{n,m}$ independent of the sign of $\xi$, allows to recover the norm as the distance in the $\xi\to 0$ limit,  
\begin{equation}
\lim_{\xi\to 0} d_{n,m}= |n-m|,
\end{equation}
and sets the unit-length distance equal to $d_{0,1}$.

\begin{figure}[htbp]
\hfill \includegraphics[width=0.8\columnwidth,keepaspectratio=true]{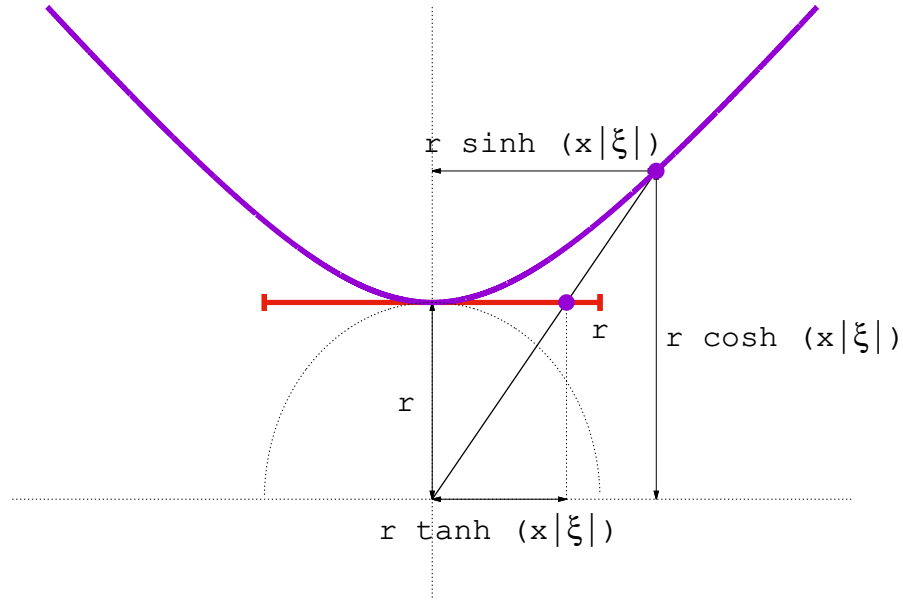}
\caption{
Cayley-Klein hyperbolic geometry. In this case, the segment in red corresponds to the absolute of the Beltrami-Klein model in one dimension: Any point of the hyperbola maps into a point in the segment. 
} 
\label{Fig:geometry}
\end{figure}

In order to clarify the meaning of equation~\eqref{eq:prob_distance} we will extend the definition in equation~\eqref{eq:dist} to any pair of points in $\RR$, 
\begin{equation}
d_{x,y}\equiv \frac{\sinh(|x-y| \xi)}{\tanh (\xi) \cosh (x\xi) \cosh (y\xi)},
\label{eq:dist_gen}
\end{equation}
with
\begin{equation}
d_{x,z}\leq d_{x,y}+d_{y,z},
\end{equation}
and where the equality holds if and only if $y \in [x,z]$. Equation~\eqref{eq:dist_gen} determines a distance in the absolute of a one-dimensional hyperbolic geometry: Consider $u$ and $v$ in the segment $(-r,r)$, with the following Cayley-Klein metric defined in it, 
\begin{equation}
\delta(u,v)\equiv\frac{1}{2}\mathrm{acosh}\left(1+\frac{2 (u-v)^2}{(r^2-u^2)(r^2-v^2)}\right).
\end{equation}
Then, if we assume the Beltrami-Klein projection of the hyperbolic geometry, see figure~\ref{Fig:geometry},
we will have
\begin{equation*}
u(x)=r \tanh (x |\xi|),\quad v(y)=r \tanh (y |\xi|),
\end{equation*}
with
\begin{equation}
r=\frac{1}{|\tanh (\xi)|}.
\end{equation}
This mapping leads to
\begin{equation}
d_{x,y}=|u(x)-v(y)|,
\label{eq:d_dist}
\end{equation}
and conversely
\begin{equation}
\delta(u,v)=|x-y|\cdot|\xi|.
\label{eq:delta_dist}
\end{equation}

\begin{figure}[htbp]
{
\hfill \includegraphics[width=0.8\columnwidth,keepaspectratio=true]{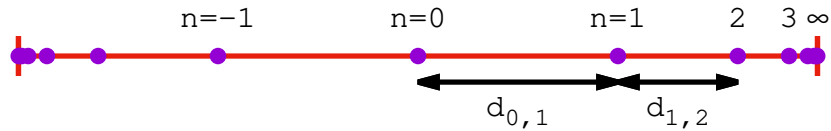}
}
\caption{
Distances in the projective segment. Here we show Euclidean distances of equally spaced points in the hyperbolic geometry. Note how the sites are closer as $|n|$ increases what biases the walk towards the ending points.} 
\label{Fig:points}
\end{figure}

Therefore, $|\xi|$ plays the role of unit of length in the hyperbolic geometry, where we have placed the equally spaced sites among which our inhomogeneous random walk evolves, see equation~\eqref{eq:delta_dist}. The probabilities, however, do not stem from the distances between the points in the hyperbolic geometry, but from the associated Euclidean distances in the absolute, the projective segment $(-r,r)$, equation~\eqref{eq:d_dist}.  We show in figure~\ref{Fig:points} a set of points uniformly sampled in terms of the hyperbolic measure $\delta$, and spatially arranged according to the Euclidean distance between them, $d_{n,n+1}$. Observe how this distance decreases as $|n|$ increases, since the Euclidean length of the segment is finite. 

\section{Conclusions and future work\label{Sec:conclusions}}

The random walk with hyperbolic probabilities that we have introduced here is an interesting example of stochastic diffusion through a one-dimensional inhomogeneous discrete media. In some aspects, it behaves as the (random) superposition of two biased random walks, but preserving traits and relationships of a simple random walk. The reason behind this latter point can be traced back to a remarkable feature of the process: even the one-step transition probabilities are site dependent, the probability of performing a closed loop of any size is not. This fact has strong implications when deriving the statistical properties related to first-time events. 

A magnitude connected with one of these first-time events is the likelihood of eventually reaching a target, if the process is presently at some given location. We have shown that the transient character of the process makes this probability always less that one, and how this leads to a 
restricted form of non-ergodicity: On the one hand, it is unlikely that  a single path is representative of all the properties of the model, even when the observation time tends to infinity, but, on the other hand, some observables can be accurately estimated from any realization of the process.

Finally, we provide an interpretation of our site-dependent transition probabilities by resorting to geometrical arguments. We leave for a future work the search for alternative models based on the interaction of a discrete medium with an external potential that can account for the emergence of these same hyperbolic transitions rates, the extension of the dynamics to geometries of higher dimensions, and the study of how the process can be transformed from the discrete to the continuum, and thus compared with similar existing models \cite{CM96,MT96,CMY98}.

\ack
The author thanks the anonymous referees for their enlightening comments that helped to considerably improve the paper. He also acknowledges support from the Spanish Agencia Estatal de Investigaci\'on and from the Fondo Europeo de Desarrollo Regional (AEI/FEDER, UE) under Contract No. FIS2016-78904-C3-2-P, and from the Catalan Ag\`encia de Gesti\'o d'Ajuts Universitaris i de Recerca (AGAUR), Contract No. 2017SGR1064. 

\end{document}